\documentclass[onecolumn,preprintnumbers,amsmath,amssymb,aps,prd,nofootinbib]{revtex4}

\usepackage{graphicx,epsfig}

\usepackage{bm}

\usepackage{amsmath}

\usepackage{hyperref}

\def\be{\begin{equation}}
\def\ee{\end{equation}}
\def\beq{\begin{eqnarray}}
\def\eeq{\end{eqnarray}}
\def\ie{{\it i.e.,}\ }
\def\eg{{\it e.g.,}\ }
\def\jmart{{JMaRT}\ }
\def\ssc{\scriptscriptstyle}

\newcommand{\reef}[1]{(\ref{#1})}

\begin{document}

\centerline{}


\title{On the gravitational stability of D1-D5-P black holes}

\author{Vitor Cardoso}
\email{vcardoso@phy.olemiss.edu} \affiliation{Department of Physics
and Astronomy, The University of Mississippi, University, MS
38677-1848, USA\footnote{Also at: Centro de F\'{\i}sica
Computacional, Universidade de Coimbra, P-3004-516 Coimbra,
Portugal}}
 \author{\'Oscar J. C. Dias}
 \email{odias@ub.edu}
\affiliation{ Departament de
F\'{\i}sica Fonamental, Universitat de Barcelona, 
Av. Diagonal 647, E-08028 Barcelona, Spain}
 \author{Robert C. Myers}
\email{rmyers@perimeterinstitute.ca}
\affiliation{Perimeter
 Institute for Theoretical Physics,
 Waterloo, Ontario N2L 2Y5, Canada;}
\affiliation{ Department of Physics and Astronomy, University of
Waterloo, Waterloo, Ontario N2L 3G1, Canada;}
\affiliation{ Kavli
Institute for Theoretical Physics, University of California, Santa
Barbara, CA 93106-4030, USA}


\begin{abstract}
We examine the stability of the nonextremal D1-D5-P black hole
solutions. In particular, we look for the appearance of a
superradiant instability for the spinning black holes but we find no
evidence of such an instability. We compare this situation with that
for the smooth soliton geometries, which were recently observed to
suffer from an ergoregion instability, and consider the implications
for the fuzzball proposal.
\end{abstract}

\maketitle

\section{\label{sec:Introduction}Introduction}

Despite being well-understood classically, black holes still pose a
number of unanswered questions at the quantum level, such as the
information paradox. A radical new approach to describe stringy
black holes, now known as the `fuzzball' proposal, was first
suggested some years ago by Mathur and collaborators \cite{fuzzy}.
They advocated that the microstates underlying a black hole are
individually described by horizon-free geometries and that the black
hole geometry only emerges in a coarse-grained description which
`averages' over the $e^{S_{\rm BH}}$ microstate geometries. In this
approach, the effective horizon of a black hole appears as a surface
at a radius where the individual microstate geometries start to
`differ appreciably' from one another. Therefore quantum gravity
effects are not confined close to the black hole singularity but
rather the entire interior of the black hole is `filled' by
fluctuating geometries -- hence the nomenclature: the `fuzzball'
description of black holes.

Finding evidence to support this conjecture is not easy though,
especially because it entails finding families of horizon-free
geometries sufficiently extensive to describe all of the $e^{S_{\rm
BH}}$ microstates for a given black hole. Most of the studies to
date focus on supersymmetric configurations, namely the BPS D1-D5
system \cite{two}, the BPS D1-D5-P system \cite{three12,three2}, and
the BPS D1-D5-P-KK system \cite{four}. However, if the fuzzball
proposal is to be useful, it must also be extended to
non-supersymmetric systems. This then poses the extremely difficult
problem of finding non-supersymmetric (smooth) horizon-free
geometries that can be associated with the microstates of a non-BPS
black hole. So far, the only known solutions in this class are those
of Jejjala, Madden, Ross and Titchener \cite{ross}, hereafter
referred to as \jmart solitons. The \jmart solutions comprise a
five-parameter family of D1-D5-P non-supersymmetric smooth
geometries which are asymptotically flat. These solutions may be
parameterized by the D1-brane and D5-brane charges, the (asymptotic)
radius of the internal circle with Kaluza-Klein momentum, and by two
integers $m$ and $n$ which fix the remaining physical parameters.
These integers also determine a spectral flow in the CFT which
allows the underlying microstate to be identified. For $m=n+1$, the
\jmart solitons reduce to supersymmetric solutions found previously
in \cite{two,three12}. The geometry of these solutions was also
recently examined in more detail in \cite{newross}.

In a previous paper \cite{Cardoso:2005gj}, we have shown that the
non-supersymmetric \jmart solitons are classically unstable against
an ergoregion instability \cite{friedman}. This kind of instability
is generic to geometries with an ergoregion but no horizon, as was
first noticed in \cite{friedman}. Thus this instability should be a
robust feature of any smooth horizon-free geometry corresponding to
a non-BPS microstate with angular momentum. At first sight then,
this seems to pose a challenge for the fuzzball proposal: if
horizon-free non-supersymmetric solitons are expected to describe
the microstates of non-BPS black holes, then the nonextremal D1-D5-P
black hole which results from averaging over an ensemble of such
solutions should be unstable against an analogous instability. The
purpose of the present work is to address this issue. While the
presence of an event horizon eliminates the possibility of an
ergoregion instability, there is, however, an obvious candidate
instability in the black hole case: the superradiant instability
\cite{Cardoso:2005gj,super1,super2,super3}.

In general, spinning nonextremal black holes will exhibit
superradiant scattering, whereby an incident wave packet can be
reflected with a stronger amplitude. Superradiance by itself does
not provide a classical instability, but an instability can arise if
the waves are reflected back and forth. Stated in other words, a
superradiant instability is present if there are bound states
subjected to superradiance. Some examples of black hole systems
unstable against such mechanism are: i) the black hole bomb where an
artificial mirror surrounds a Kerr black hole \cite{press}; ii) a
massive scalar field in a Kerr background \cite{detweiler}; iii)
small Kerr-AdS black holes \cite{CardDiasAdS}; iv) rotating black
strings \cite{super1,super2,super3}. This instability seems to be
the natural extension, to black holes, of the ergoregion instability
found in \cite{Cardoso:2005gj}. Therefore, according to the fuzzball
proposal, one might expect that the black hole family of D1-D5-P
non-supersymmetric geometries should be superradiantly unstable --
however, we find they are not! Nevertheless, as we will discuss,
this does not present a sharp contradiction with the fuzzball
proposal. 

The remainder of this paper is organized as follows: In Section
\ref{sec:properties metric}, we briefly review some of the relevant
properties of  the D1-D5-P family of supergravity solutions. In
Sections \ref{sec:WaveSep} and \ref{sec:Sch}, we study minimally
coupled scalar waves in the general D1-D5-P background. We write the
wave equation in a Schr\"{o}dinger form and perform an extensive
search for unstable free scalar modes in the D1-D5-P black hole
geometry. We find no unstable modes. The same conclusion applies for
a massive scalar field. In Section \ref{sec:Conc}, we discuss our
results and consider some of their implications for the fuzzball
proposal. In particular, we discuss possible ways the fuzzball proposal can 
consistently incorporate our results. The paper contains two
appendices: In Appendix \ref{sec:JMaRT}, we show that the wave
equation for the general D1-D5-P system written in Section
\ref{sec:WaveSep} reduces to the form presented in
\cite{ross,Cardoso:2005gj} when we restrict our analysis to the
\jmart solitons. In Appendix \ref{sec:potential}, we give the
explicit expression for the Schr\"{o}dinger potentials introduced in
section \ref{sec:Sch}.

\section{\label{sec:properties metric}Properties of the D1-D5-P family of solutions}

The black hole and solitonic configurations  considered here are
solutions of the type IIb supergravity equations. The solutions all
carry the three charges of the D1-D5-P system and so are expected to
give a strong-coupling description of different (ensembles of)
microstates of this system. The supergravity solution is comprised
of a (ten-dimensional) metric and also a nontrivial dilaton and RR
two-form potential. The system is compactified to five dimensions on
$M^4 \times S^1$ with the D5-branes wrapping the full internal space
and the D1-branes and KK-momentum on the distinguished $S^1$. The
other component of the compactification $M^4$ is a Ricci-flat
fouf-manifold, which we take to be either a four-torus $T^4$ or
$K3$. The notation is best understood by considering the
construction of these solutions. One begins with the general
solutions of \cite{BLMPSV,CY,three12} which contain eight
parameters: a mass parameter, $M$; spin parameters in two orthogonal
planes, $a_1,a_2$ (which we assume are non-negative without loss of
generality); three boost parameters, $\delta_1,\delta_5,\delta_p$,
which fix the D1-brane,\footnote{In the following, only the
asymptotic D1-brane charge $Q_1$ appearing in the supergravity
fields will be relevant. Of course, with $M^4=K3$, $Q_1\propto
N_1-N_5$ where $N_1$ and $N_5$ are the numbers of constituent D1-
and D5-branes, respectively, comprising the system \cite{Kthree}.
Ref.~\cite{cool} describes how this technical point produces an
interesting physical effect for the D1-D5-P black holes considered
here.} D5-brane and KK-momentum charges, respectively; the radius of
the $S^1$, $R$; the volume of the four-manifold, $V_4$ (which plays
no role in the following). The geometry is described by the
six-dimensional line element which is given below (see also Equation
(2.12) of \cite{ross}) and which is parameterized by a time
coordinate $t$; a radial coordinate $r$; three angular coordinates
$\theta$, $\phi$, $\psi$; and the coordinate on the $S^1$, $y$.

The (ten-dimensional) dilaton and two-form RR gauge potential which
support the D1-D5-P configuration are \cite{three12,ross}
\begin{eqnarray}
&& e^{2\Phi_{\ssc 10}} = {\tilde H_1}/{\tilde H_5}\,, \label{dilatone} \\
&& C^{(2)} = \frac{M \cos^2 \theta}{\tilde H_1} \left[ (a_2 c_1
  s_5 c_p - a_1 s_1 c_5
  s_p) dt + (a_1 s_1 c_5 c_p - a_2 c_1 s_5 s_p) dy \right] \wedge d\psi
   \nonumber \\
&& \hspace{1cm}+ \frac{M \sin^2 \theta}{\tilde H_1} \left[  (a_1 c_1
  s_5 c_p - a_2 s_1 c_5
  s_p) dt  + (a_2 s_1 c_5 c_p - a_1 c_1 s_5 s_p) dy \right] \wedge d \phi
  \nonumber \\
&& \hspace{1cm} - \frac{M s_1 c_1}{\tilde H_1} dt \wedge dy -
  \frac{M s_5 c_5}{\tilde H_1} (r^2 + a_2^2 + M
  s_1^2) \cos^2 \theta d\psi \wedge d\phi. \label{rr}
\end{eqnarray}
However, these will only play an ancillary role in the present
discussion. Central to our analysis will be the six-dimensional
geometry consisting of the noncompact space, as well as the
$y$-circle. The contravariant components of the string-frame metric
are\footnote{The boosted coordinates $(\tilde t, \tilde y)$ are
related to the unboosted coordinates $(t,y)$ by $\tilde t = t c_p -
y s_p$, $\tilde y = y c_p -t s_p $ -- see Appendix A of \cite{ross}.
\label{footer}}
\begin{align}
g^{tt}&= c_p^2 g^{\tilde{t}\tilde{t}} + 2s_p c_p g^{\tilde{t}\tilde{y}}
       +s_p^2 g^{\tilde{y}\tilde{y}}\,, \qquad
g^{ty}= s_p c_p (g^{\tilde{t}\tilde{t}}+ g^{\tilde{y}\tilde{y}})+
       (c_p^2+s_p^2) g^{\tilde{t}\tilde{y}} \,, \qquad
g^{yy}= s_p^2 g^{\tilde{t}\tilde{t}} + 2s_p c_p
       g^{\tilde{t}\tilde{y}}+c_p^2 g^{\tilde{y}\tilde{y}}\,,
\nonumber \\
g^{t\phi}&= c_p g^{\tilde{t}\phi} +s_p g^{\tilde{y}\phi} \,,\qquad
g^{t\psi}= c_p g^{\tilde{t}\psi} +s_p g^{\tilde{y}\psi}\,,\qquad
g^{y\phi}= s_p g^{\tilde{t}\phi} +c_p g^{\tilde{y}\phi}\,,\qquad
g^{y\psi}= s_p g^{\tilde{t}\psi} +c_p g^{\tilde{y}\psi}\,, \nonumber
\end{align}
\begin{align}
g^{rr} &= \frac{1}{\sqrt{\tilde H_1 \tilde H_5}} \frac{g(r)}{r^2}
\,,\qquad
 g^{\theta \theta} = \frac{1}{\sqrt{\tilde H_1 \tilde H_5}}
\,,\qquad
 g^{\phi\phi} = \frac{1}{\sqrt{\tilde H_1 \tilde H_5}} \left(
  \frac{1}{\sin^2 \theta} + \frac{(r^2+a_1^2)(a_1^2-a_2^2)-M a_1^2}{g(r)} \right) , \nonumber \\
g^{\phi\psi} &= -\frac{1}{\sqrt{\tilde H_1 \tilde H_5}} \frac{Ma_1
  a_2}{g(r)}\,,\qquad
g^{\psi\psi} = \frac{1}{\sqrt{\tilde H_1 \tilde H_5}}\left(
  \frac{1}{\cos^2 \theta}
  + \frac{(r^2+a_2^2)(a_2^2-a_1^2)-M a_2^2}{g(r)} \right),
  \label{stringmetric}
\end{align}
where\footnote{Eqs. (A.2), (A.3) and (A.4) of \cite{ross} have a
typo and must be multiplied by minus one.}
\begin{align}
g^{\tilde t\tilde t} &= - \frac{1}{\sqrt{\tilde H_1 \tilde H_5}}
  \left( f(r) + M + M s_1^2+ M s_5^2 + \frac{M^2 c_1^2 c_5^2 r^2}{g(r)} \right), \nonumber \\
g^{\tilde t \tilde y} &= -\frac{1}{\sqrt{\tilde H_1 \tilde H_5}}
\frac{M^2 s_1
  s_5 c_1 c_5 a_1 a_2}{g(r)},\qquad
g^{\tilde t\phi} = \frac{1}{\sqrt{\tilde H_1 \tilde H_5}} \frac{M
c_1
  c_5 a_2 (r^2+a_1^2)}{g(r)},\qquad
g^{\tilde t\psi} = \frac{1}{\sqrt{\tilde H_1 \tilde H_5}} \frac{M
c_1
  c_5 a_1 (r^2+a_2^2)}{g(r)}, \nonumber \\
g^{\tilde y \tilde y} &= \frac{1}{\sqrt{\tilde H_1 \tilde H_5}}
\left(
  f(r) + M s_1^2
  + M s_5^2 + \frac{M^2 s_1^2 s_5^2 (r^2 +
  a_1^2 + a_2^2 -M)}{g(r)} \right) , \nonumber \\
g^{\tilde y\phi} &= -\frac{1}{\sqrt{\tilde H_1 \tilde H_5}}\frac{M
s_1 s_5 a_1 (r^2+a_1^2-M)}{g(r)},\qquad
 g^{\tilde y\psi} =
-\frac{1}{\sqrt{\tilde H_1 \tilde H_5}} \frac{M s_1 s_5 a_2
(r^2+a_2^2-M)}{g(r)}\,.
\end{align}
We are using the notation $c_i \equiv \cosh \delta_i$ and $s_i
\equiv \sinh \delta_i$. Throughout these expressions, we also use
the functions:
\begin{eqnarray}
 g(r)&=&(r^2+a_1^2)(r^2+a_2^2)-M r^2\,\,
\equiv  (r^2-r_+^2)(r^2-r_-^2) \label{def function g} \,,\\
f(r)&=&r^2+a_1^2\sin^2\theta+a_2^2\cos^2\theta \,, \qquad
\tilde{H}_i(r)=f(r)+M s_i^2\,, \:{\rm with}\: i=1,5\,.
 \label{def f H}
 \end{eqnarray}
Without loss of generality, we will assume $a_1 \ge a_2\ge 0$ in the
following.

Depending on the values of the parameters, this solution can
describe a black hole or a naked curvature singularity. However, it
was also realized that in a third parameter regime \cite{ross}, this
geometry corresponds to a smooth soliton -- denoted the \jmart
soliton in \cite{Cardoso:2005gj} -- or a conical singularity. The
best way to identify each one of these branches of solutions is to
look at the $g^{rr}$ component of the general metric, which is
proportional to the function $g(r)$, given in
 eq.~\reef{def function g} above. The roots of $g(r)$, $r_+$ and
 $r_-$, are given by \cite{ross}
\begin{eqnarray}
 r_{\pm}^2= \frac{1}{2}\,(M-a_1^2-a_2^2) \pm \frac{1}{2}
 \sqrt{(M-a_1^2-a_2^2)^2-4a_1^2 a_2^2}\,, \label{r+-}
\end{eqnarray}
and they are real whenever $|M-a_1^2-a_2^2|>2a_1a_2$. We can
naturally divide the general family of solutions into three
branches, namely \cite{ross}:
\begin{equation}
 \left\{
\begin{array}{lll}
 M \leq(a_1-a_2)^2  \qquad \qquad \qquad \qquad \Rightarrow  r_+^2<0 \,, \qquad
 ({\rm \jmart}\: {\rm branch}) \,, \\
 (a_1-a_2)^2< M< (a_1+a_2)^2  \qquad  \: \Rightarrow  r_+^2 \notin \mathbb{R} \,,
  \qquad ({\rm Naked}\: {\rm singularity}\: {\rm branch}) \,, \\
M\geq (a_1+a_2)^2 \qquad \qquad \qquad \qquad \Rightarrow  r_+^2>0
\,, \qquad ({\rm Black} \: {\rm hole} \: {\rm branch})\,.
\end{array}
\right.
\label{branches}
\end{equation}
In the first case the system can describe either a smooth soliton or
a conical singularity. (In Appendix \ref{sec:JMaRT}, we present the
constraints which must be satisfied to produce a smooth geometry.)
In the second case, the function $g(r)$ does not have real roots and
the system describes a naked singularity, with curvature
singularities where $\tilde{H}_i(r)$ vanishes. Finally, in the third
case, the system describes a black hole with outer horizon at
$r^2=r_+^2$ and inner horizon at $r^2=r_-^2$. The upper bound of the
first branch, $M^2=(a_1-a_2)^2$ (for which $r_+^2=r_-^2=-a_1 a_2$),
includes as special cases the full set of supersymmetric smooth
solitons \cite{two,three12}. The lower bound of the third branch, $M
= (a_1+a_2)^2$ , corresponds to an extremal black hole with
$r_+^2=r_-^2=a_1 a_2$.  The supersymmetric limit of the above
three-charge system corresponds to taking the limit $M\rightarrow 0$
and $\delta_i \rightarrow \infty$, while keeping the other
parameters fixed, including the conserved charges $Q_i=M s_i c_i$.
One gets, in the non-singular case, the black hole solutions of
\cite{Breckenridge:1996is} or the supersymmetric solitons of
\cite{two,three12} -- see \cite{ross}.

In this paper, the key feature of interest in this geometry is the
ergosphere. In particular, we will be considering instabilities that
arise due to the existence of an ergoregion. To verify the presence
of the ergoregion, one may take the norm of the Killing vector
$V=\partial_t$ yielding
\begin{eqnarray}
 g_{AB}V^{A}V^{B} =-\frac{f-Mc_p^2}{\sqrt{\tilde{H}_1
 \tilde{H}_5}}\,.
 \label{ergoregionV}
\end{eqnarray}
This result shows that $V=\partial_t$ becomes space-like for
$f(r)<Mc_p^2$ and thus one would conclude that an ergosphere appears
at $f(r)=Mc_p^2$. However, in the supersymmetric limit, this norm
\reef{ergoregionV} becomes $|V|^2=-(f-Q_p)/\sqrt{\tilde{H}_1
\tilde{H}_5}$. So using this measure, one arrives at the
counter-intuitive conclusion that the ergoregion persists even for
the supersymmetric backgrounds. The resolution of this puzzle is
evident in the discussion of section 6.2 of \cite{ross}. The key
point is that the present geometry `rotates' along the internal
$y$-direction, as well as along the angles $\phi$ and $\psi$. Hence
for the purposes of defining the ergosphere, we have a continuous
family of asymptotically time-like Killing vectors:
$\widetilde{V}=\partial_t + v^y \partial_y$ with
$|v^y|<1$.\footnote{We note that $|v^y|=1$ would yield an
asymptotically null Killing vector. Of course, the norm of the
angular Killing vectors $\partial_\phi$ and $\partial_\psi$ diverges
asymptotically and so we cannot consider further linear combinations
by including either of these vectors.} Now one can push the position
of the ergosphere to smaller radii by adjusting the free parameter
$v^y$. In particular, the position seems to be minimized by the
choice $v^y=\tanh\, \delta_p$, for which
$\widetilde{V}=\partial_{\tilde t}$ (\ie we have undone the boost of
footnote \ref{footer}) and
\begin{eqnarray}
 g_{AB}\widetilde{V}^{A}\widetilde{V}^{B}
 =-\frac{f-M}{\sqrt{\tilde{H}_1\tilde{H}_5}}\,.
 \label{ergoregionVtil}
\end{eqnarray}
Further we find that this choice of $\widetilde{V}$ matches the
Killing vector arising from the square of the covariantly constant
Killing spinor appearing in the BPS limit. Of course, the ergoregion
now disappears in this supersymmetric limit where $M\rightarrow0$
and $f\rightarrow r^2$. A perspective relevant for the following
discussion is that in the supersymmetric backgrounds,
$\widetilde{V}$ provides a globally time-like Killing vector which
ensures that there is a 'rotating' or `boosted' frame where all
energies can be defined to be positive.

\section{\label{sec:WaveSep}Separation of the wave equation in the general D1-D5-P system}
The \jmart solitons were found to be classically unstable in
ref.~\cite{Cardoso:2005gj}. What we wish to do now is apply a
similar stability analysis to three-charged black hole geometry to
ascertain whether or not these solutions are also unstable. A
stability analysis starts by disturbing slightly the system and then
letting it evolve freely. Stable systems return to their original
`position' whereas in unstable ones, the perturbations will grow
without bound in time. Ideally one would like to consider
perturbations by any of the type IIb supergravity fields, \eg
metric, dilaton or RR two-form perturbations. In Sections
\ref{sec:WaveSep} and \ref{sec:Sch}, we will begin by considering
only free scalar perturbations that obey the Klein-Gordon equation.
These perturbations are expected to capture the essential physics
while at the same time simplifying enormously the calculations.
However, as we discuss in Section \ref{sec:Conc}, our results for
the free scalar apply directly to certain supergravity fields.

So we first consider the Klein-Gordon equation for a massless scalar
field propagating in a general three-charge geometry. The results of
this section are valid for the full range of the parameters defining
the configuration space. Let us note that our strategy will be to
solve the wave equation in the six-dimensional background formed by
the non-compact directions and the distinguished $S^1$ circle. That
is, we limit our analysis so that the directions along internal
manifold $M^4$ play no role. This amounts to having a wave with no
excitations along these internal directions. 
We must mention that a related analysis of this equation in the
effective five-dimensional geometry coming from also reducing on the
$S^1$ circle first appeared in \cite{finn}.\footnote{The present
analysis is made in six, rather than five, dimensions because the
Kaluza-Klein momentum along the $S^1$ plays an important role, as
discussed at the end of that section \ref{sec:Sch}. Further the
coordinates in \cite{finn} are adapted to calculating the absorption
cross-section of the black hole rather than producing the
Schr\"odinger form of section \ref{sec:Sch}.}

We must precede our analysis of the wave equation with the
observation that for the D1-D5-P backgrounds, the effective
six-dimensional dilaton vanishes: The ten-dimensional dilaton is
given in eq.~\reef{dilatone} while the string-frame metric on the
internal space $M^4$ takes the form
 \beq
 {G}_{ab}=\left( \frac{\tilde{H}_1}{\tilde{H}_5}
\right)^{\frac{1}{2}} {\widetilde G}_{ab} = e^{\Phi_{\ssc 10}}\,
{\widetilde G}_{ab}\,,
 \label{eta}
  \eeq
where ${\widetilde G}_{ab}$ is a fiducial metric with unit
four-volume. Hence upon reducing the supergravity action to six
dimensions, we find the effective six-dimensional dilaton becomes:
\beq e^{-2\Phi_{\ssc 6}}\equiv e^{-2\Phi_{\ssc 10}}\, \sqrt{{\rm
det}{G}_{ab}}=1\,.\label{dilatonsix}\eeq
It follows then that in the effective six-dimensional theory, there
is no distinction between the string-frame and Einstein-frame
metrics. In particular, the six-dimensional metric has precisely the
same components as those of the ten-dimensional string-frame metric,
given in eq.~\reef{stringmetric}.

Hence we consider the Klein-Gordon equation in the effective
six-dimensional theory\footnote{This wave equation is equivalent to
that which would be produced in reducing the equation of motion of a
massless scalar field which is minimally coupled to the
Einstein-frame metric (and no other fields) in ten dimensions.}
coming from the reduction of the general three-charge geometry:
\begin{eqnarray}
 \frac{1}{\sqrt{-g}}\frac{\partial}{\partial x^{A}}
\left(\sqrt{-g}\,g^{AB}\frac{\partial}{\partial x^{B}}\Psi
\right)=0\,,
 \label{klein}
\end{eqnarray}
where $g^{AB}$ is the six-dimensional metric explicitly written
above, in section \ref{sec:properties metric}. Further $g$ is the
determinant of this six-dimensional metric, which is given by
\begin{eqnarray}
\sqrt{-g}=r\sin\theta\cos\theta \sqrt{{\tilde H_1}{\tilde H_5}}\,.
 \label{gString6}
\end{eqnarray}
This is the same equation studied in the stability analysis of
\cite{Cardoso:2005gj}. Introducing the separation ansatz
\begin{eqnarray}
\Psi=\exp\left[-i\omega \frac{t}{R}-i\lambda \frac{y}{R}+i
m_{\psi} \psi +i m_{\phi} \phi \right]\, \chi(\theta)\,h(r) \,,
 \label{separation ansatz}
\end{eqnarray}
and the separation constant $\Lambda$, the wave equation separates.
The angular equation is
\begin{eqnarray}
 \frac{1}{\sin{2\theta}}\frac{d}{d\theta}\left
(\sin{2\theta}\,\frac{d\chi}{d\theta}\right )+{\biggr
[}\Lambda-\frac{m_{\psi}^2}{\cos^2\theta}-\frac{m_{\phi}^2}{\sin
^2\theta}+\frac{\omega ^2-\lambda ^2}{R^2}(a_1 ^2\sin
^2\theta+a_2^2\cos^2\theta) {\biggr ]}\chi=0 \label{angeq}\, ,
\end{eqnarray}
and the radial equation is
  \begin{eqnarray}
\begin{aligned} & \frac{1}{r} \frac{d}{dr} \left[ \frac{g(r)}{r}
\frac{d}{dr} h \right]
  - \Lambda h + \left[ \frac{(\omega^2 - \lambda^2)}{R^2} (r^2 + M s_1^2 + M
    s_5^2) + (\omega c_p + \lambda s_p)^2 \frac{M}{R^2} \right] h
  \\
  & +\frac{1}{g(r)} {\biggl \{ } -\omega^2\,\frac{M^2}{R^2}
  \left[ - c_1^2 c_5^2 c_p^2  r^2
  - 2 s_1 s_5 s_p c_1 c_5 c_p  a_1 a_2
  + s_1^2 s_5^2 s_p^2  (r^2+a_1^2+a_2^2-M)\right]
  \\
  &  \hspace{1.5cm}+ 2\omega\lambda \frac{M^2}{R^2}\,\left[  -s_p c_p \left[-c_1^2
c_5^2   r^2 + s_1^2 s_5^2  (r^2+a_1^2+a_2^2-M)\right]+
(c_p^2+s_p^2) s_1 s_5 c_1 c_5  a_1 a_2\right]
 \\
  &  \hspace{1.5cm}+\frac{2\omega m_{\phi}}{R} M \left[ c_1
c_5 c_p  a_2 (r^2+a_1^2) - s_1 s_5 s_p  a_1 (r^2+a_1^2-M) \right]
  \\
  & \hspace{1.5cm}+\frac{2\omega m_{\psi}}{R} M \left[ c_1
c_5 c_p a_1 (r^2+a_2^2) - s_1 s_5 s_p  a_2 (r^2+a_2^2-M) \right]
   \\
  &  \hspace{1.5cm}-\lambda^2 \,\frac{M^2}{R^2} \left[ - c_1^2 c_5^2 s_p^2  r^2
  - 2 s_1 s_5 s_p c_1 c_5 c_p  a_1 a_2
  + s_1^2 s_5^2 c_p^2  (r^2+a_1^2+a_2^2-M) \right]
     \\
  & \hspace{1.5cm}+\frac{2\lambda m_{\phi}}{R} M \left[ c_1
c_5 s_p  a_2 (r^2+a_1^2) - s_1 s_5 c_p  a_1 (r^2+a_1^2-M) \right]
  \\
  &  \hspace{1.5cm}+\frac{2\lambda m_{\psi}}{R} M \left[ c_1
c_5 s_p  a_1 (r^2+a_2^2) - s_1 s_5 c_p  a_2 (r^2+a_2^2-M) \right]
  \\
  &  \hspace{1.5cm}- m_{\phi}^2 \left[ (r^2+a_1^2)(a_1^2-a_2^2)-M a_1^2\right]
- m_{\psi}^2 \left[(r^2+a_2^2)(a_2^2-a_1^2)-M a_2^2\right] +2m_{\phi}m_{\psi}
  M a_1 a_2 {\biggr \}}h
    = 0 \label{radialeq-r}\,,
\end{aligned}
\end{eqnarray}
This radial wave equation is valid for the the general three-charge
geometries, including the black hole solutions and the \jmart
solitons. However, as shown in Appendix \ref{sec:JMaRT}, when the
\jmart constraints
(\ref{JMaRTconstraints})-(\ref{JMaRTconstraints2}) are imposed, the
wave equation can be rewritten in a considerably simplified way,
namely as in eq.~(\ref{WaveEqJMaRT}).

The angular equation (\ref{angeq}), plus the appropriate regularity
requirements, defines a Sturm-Liouville problem, and the solutions
are known as higher dimensional spin-weighted spheroidal harmonics
\cite{Berti:2005gp}. We can label the corresponding eigenvalues
$\Lambda$ with an index $l$, $\Lambda(\omega)=\Lambda_ {l m_\phi
m_\psi}(\omega)$ and therefore the wavefunctions form a complete set
over the integer $l$. In the general case, the problem at hand
consists of two coupled second-order differential equations: given
some boundary conditions, one has to compute simultaneously both
values of $\omega$ and $\Lambda$ that satisfy these boundary
conditions. However, for vanishing $a_i ^2$ we get the
(five-dimensional) flat space result, $\Lambda=l(l+2)$, and the
associated angular functions are simply given by Jacobi polynomials
\cite{Berti:2005gp}. For non-zero, but  small $\frac{\omega
^2-\lambda ^2}{R^2}a_i ^2$ we have
\begin{eqnarray} \Lambda=l(l+2)+\mathcal{O}
\left (a_i^2\frac{\omega^2-\lambda ^2}{R^2}\right) \label{app} \,.
\end{eqnarray}
The integer $l$ is constrained to be $l\geq |m_{\psi}|+|m_{\phi}|$.
We will always assume $a_i^2\frac{\omega^2-\lambda ^2}{R^2} \ll {\rm
max}(m_{\psi}^2,m_{\phi}^2)$ (with $i=1,2$) and thus $\Lambda \simeq
l(l+2)$. Making this assumption implies we may neglect the terms
proportional to $a_i$ in the angular equation. But given the way
$\Lambda$ and $\omega$ appear in the radial equation, the
corrections to $\Lambda$ may not be negligible when we determine
$\omega$.  To ensure that setting $\Lambda=l(l+2)$ is consistent in
both the angular and radial equations, we must additionally require
[see first line of (\ref{radialeq-r})]: $a_i^2 \ll \max \left (
|r_+^2+ M(s_1^2+s_5^2) | , M c_p^2 \right )$.

\section{\label{sec:Sch}Wave equation in the Schr\"{o}dinger form}

In \cite{Cardoso:2005gj}, we have shown that the non-supersymmetric
\jmart solitons are unstable against the ergoregion instability. The
ingredients for this instability are the existence of an ergoregion
in a geometry without horizon. If there is a counterpart of the
ergoregion instability on the black hole side of the configuration
space, it seems likely to be the so-called superradiant instability.
The ingredients for this instability are the existence of an
ergoregion around a horizon and the presence of `bound' states
within the superradiant regime
\cite{press,detweiler,CardDiasAdS,super1,super2,super3}.

The simplest way to find if the non-supersymmetric black hole
solutions of the D1-D5-P system are superradiantly unstable is to
rewrite the radial wave equation in the form of an effective
Schr\"{o}dinger equation, and to study the corresponding
Schr\"{o}dinger potentials. We now apply this approach in the
following.

In \cite{Cardoso:2005gj} we found that the introduction of
dimensionless coordinate
\begin{eqnarray}
 x=\frac{r^2 - r_+^2}{r_+^2 - r_-^2}\,, \label{coord x}
\end{eqnarray}
along with a new wavefunction $H$
\begin{eqnarray}
 h(x)=\frac{1}{\sqrt{x(1+x)}}\,H(x)\,.
\end{eqnarray}
transformed the radial wave equation (\ref{radialeq-r}) to
\begin{eqnarray}
\partial ^2 _x H + m_{\psi}^2\frac{\cal P}{4x^2(1+x)^2}
(\Sigma_{\psi}-U_+)(\Sigma_{\psi}-U_-) \,H=0 \label{Schrod 1}\,,
\end{eqnarray}
where ${\cal P}$, $U_-$ and $U_+$ are presented in
\cite{Cardoso:2005gj}.  However, we find that this form is no
longer appropriate to study the black hole sector, since in this
case both $U_-$ and $U_+$ take on complex values, sufficiently
close to the black hole horizon $r_+$.

\begin{figure}[h]
\begin{center}
\resizebox{14cm}{6cm}
 {\includegraphics{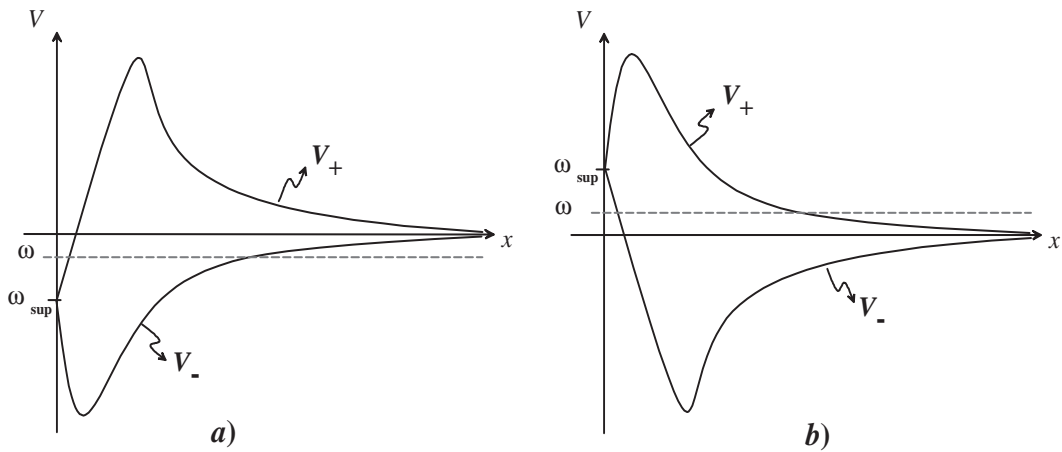}}
\end{center}
\caption{Typical Schr\"{o}dinger potentials for the black hole
branch,   $M\geq (a_1+a_2)^2$ or $r_+^2>0$, and for modes with no
KK momentum, $\lambda=0$. Figure a) refers to the case
$m_{\psi}>0$, while figure b) corresponds to case $m_{\psi}<0$.
Superradiant modes (dashed line) with $|\omega|< |\omega_{\rm
sup}|$ do exist but there are no trapped bound states and thus the
geometry is not afflicted by the superradiant instability. }
 \label{fig:bh}
\end{figure}

Instead, we introduce the `tortoise' coordinate\footnote{Note that
this coordinate is not appropriate to analyze the $M\leq
(a_1-a_2)^2$ branch of solutions since $r_+^2<0$ in this case, and
thus $r_*$ becomes complex near the origin $r^2=r_+^2$.}
\begin{eqnarray}
 \frac{dr_*}{dr}=\frac{r^4}{g(r)}\,,
\end{eqnarray}
where $g(r)$ was defined in eq.~\reef{def function g} and the new
wavefunction $\Phi$ is given by
\begin{eqnarray}
h=r^{-3/2}\Phi\,. \label{radial0}
\end{eqnarray}
Then (\ref{radialeq-r}) can be written as a Schr\"{o}dinger
equation,
\begin{eqnarray}
\frac{d^2\Phi}{dr_*^2}-V\,\Phi=0 \,,\label{radial1}
\end{eqnarray}
with
\begin{eqnarray}
V=-\frac{g(r)}{r^{10}}\left (r^4W(r)-\frac{3}{2}r
g'(r)-\frac{21}{4}g(r)\right ) \,.\label{radial2}
\end{eqnarray}
Here the prime denotes a derivative with respect to $r$, and $W(r)$
is defined by writing (\ref{radialeq-r}) in the form
$\frac{1}{r}\partial_r\left[\frac{g(r)}{r}\partial_r h\right
]+W(r)\,h=0$. As discussed earlier, we assume $\Lambda \approx
l(l+2)$ independent of $\omega$ and therefore $V$ is a quadratic
function of $\omega$
\begin{eqnarray}
V=-\gamma(\omega-V_+) (\omega-V_-) \,. \label{Vfactorized}
\end{eqnarray}
The explicit forms of $\gamma$ and of the Schr\"{o}dinger potentials
$V_{\pm}$ is given in Appendix \ref{sec:potential}. It can be
checked that the asymptotic behavior of the potentials is
\begin{eqnarray}
\lim_{r\rightarrow r_+} V_{\pm}&=& \omega_{\rm sup} \,, \\
\lim_{r\rightarrow \infty} V_{\pm}&=& \pm |\lambda| \,.
 \label{lim V} \end{eqnarray}
where we defined the superradiant factor,
\begin{eqnarray}  \omega_{\rm sup}= m_{\phi} \Omega_{\phi} R+ m_{\psi} \Omega_{\psi}
R - \lambda \Omega_y \,. \label{omega sup}
 \end{eqnarray}
Here, $\Omega_{\phi}$, $\Omega_{\psi}$, and $\Omega_y$ are,
respectively, the angular velocities along $\phi,\psi$ and the
velocity along the $S^1$ given by \cite{Dias:2007dj}
\begin{eqnarray} \Omega_{\phi,\psi} = -\frac{ a_{2,1} r_+^2}
 {\left( r_+^2+a_{2,1}^2 \right) \left( r_+^2c_1c_5c_p+a_1a_2s_1s_5s_p \right)}
 \,, \qquad\qquad
\Omega_y = \frac{ r_+^2 c_1c_5s_p + a_1a_2 s_1s_5c_p }
                 {r_+^2c_1c_5c_p+a_1a_2s_1s_5s_p}\,.
   \label{velocity}
 \end{eqnarray}
Typical forms of the potentials $V_{\pm}$ are displayed in
Fig.~\ref{fig:bh}. In these plots, the `allowed' regions where the
solutions have an oscillatory behavior are those where $\omega$ is
above $V_+$ or below $V_-$, \ie above or below both curves $V_\pm$.
In those intervals where $\omega$ is in between the curves of  $V_+$
and $V_-$ (forbidden regions), the solutions have a real exponential
behavior. From these plots we will infer the stability of the
system.
We impose only ingoing-wave boundary conditions near the horizon and
outgoing at infinity. This means that the asymptotic behavior of the
solutions of (\ref{radial0})-(\ref{radial1}), is
\begin{eqnarray} h(r) \sim \left\{
\begin{array}{ll}
(r^2-r_+^2)^{-i \varpi}=e^{-i \varpi \ln (r^2-r_+^2)}\,,
\qquad  {\rm as} \quad r\rightarrow r_+ \,, \\
r^{-2}\,e^{+i R^{-1}\sqrt{\omega^2-\lambda^2} \,r}\,,
 \qquad {\rm as} \quad r\rightarrow \infty \,,
\end{array} \right. \label{bound cond}
 \end{eqnarray}
where we have defined
\begin{eqnarray} \varpi = \frac{ M\left(r_+^2 c_1 c_5 c_p + a_1 a_2 s_1 s_5 s_p
\right)}{2R\, r_+(r_+^2-r_-^2)}\,
 (\omega-\omega_{\rm sup}) \,.
  \end{eqnarray}
When the frequency of the wave is such that $\varpi$ is negative,
\begin{eqnarray}
 |\omega|< |\omega_{\rm sup}| \,,
 \label{super cond}
  \end{eqnarray}
one is in the superradiant regime, and the  amplitude of any
incident wave is amplified upon scattering from the black hole.

The simplest way to appreciate this result is through the following
argument \cite{Bardeen:1972fi}:\footnote{See section II.A of
\cite{Cardoso:2005gj} for an alternative argument.} From
(\ref{separation ansatz}) and (\ref{bound cond}) one has that at the
horizon the wave solution behaves as $\Psi(t,r) {\bigl
|}_{r\rightarrow r_+}\sim e^{-i\omega t/R} e^{-i \varpi \ln
(r^2-r_+^2)}$. The phase velocity of the wave is then $v_{\rm ph}
\propto -\frac{\omega}{\varpi}$. Now, the value of this phase
velocity can be positive or negative depending on the value of
$\omega$ (when we fix the other parameters), so one might question
if the first line of (\ref{bound cond}) really describes {\it
always} an ingoing wave. What is relevant for the discussion is not
the phase velocity, but the group velocity of the waves. The
normalized group velocity, $v_{\rm gr}$, at the horizon is $v_{\rm
gr}= 2(r_+^2-r_-^2)r_+ \alpha^{-1}\, \frac{d(-\varpi)}{d \omega} =
-1$.\footnote{Note that for $M\geq (a_1+a_2)^2$ one has $r_+^2 \geq
a_1 a_2 $. Moreover, $c_i^2=1+s_i^2>s_i^2$.} This is a negative
value that signals that the near-horizon wave solution in
(\ref{bound cond}) always represents an ingoing wave {\it
independently} of the value of $\omega$, and thus we have the
correct physical boundary condition. However, note that in the
superradiant regime (\ref{super cond}), the phase velocity is
positive and so waves appear as outgoing to an inertial observer at
spatial infinity. Therefore energy is in fact being extracted from
the black hole.
\begin{figure}[h]
\begin{center}
\resizebox{14cm}{6cm}
 {\includegraphics{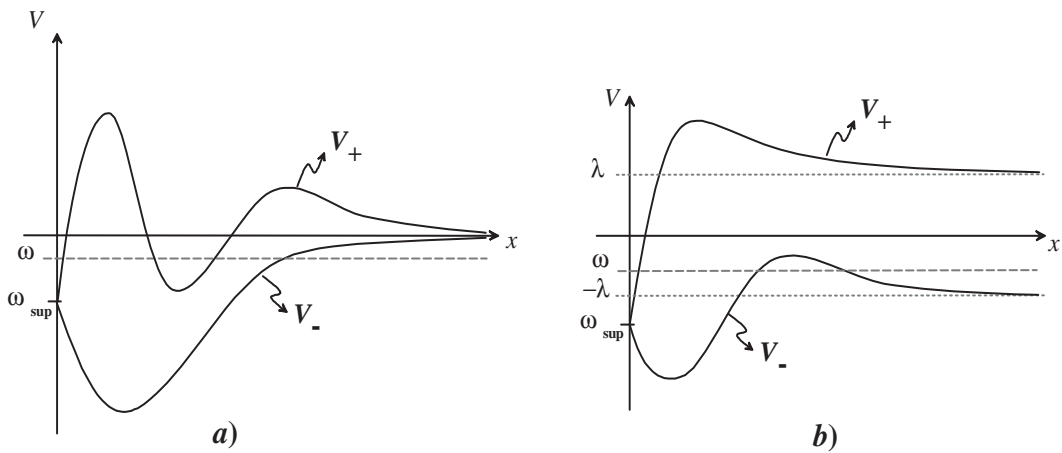}}
\end{center}
\caption{Examples of potentials that would correspond to black
hole geometries afflicted by the superradiant instability.
Unstable modes are superradiant modes with
 $|\omega|< |\omega_{\rm sup}|$ that are also bound states (dashed line).
Case $a$) corresponds to $\lambda=0$ while case $b$) corresponds
to $\lambda\neq 0$. }
 \label{fig:unstable}
\end{figure}

The relevance of superradiance in the present context is that it may
give rise to instabilities. The only extra ingredient one needs for
that to happen is to somehow `trap' the waves near the horizon. An
artificial way of doing this would be to enclose the black hole
inside a mirror \cite{press}. Any initial perturbation will get
successively amplified near the black hole event horizon and
reflected back at the mirror, thus creating an instability. This
instability is caused by the mirror, which is an artificial wall,
but one can also devise natural mirrors. For example, consider a
massive scalar field \cite{detweiler}. Imagine a wavepacket of the
massive field in a distant circular orbit. The gravitational force
binds the field and keeps it from escaping or radiating away to
infinity. But at the event horizon some of the field goes down the
black hole, and if the frequency of the field is in the superradiant
region then the field is amplified. Hence the field is amplified at
the event horizon while being confined away from infinity. Yet
another way to understand this, is to think in terms of wave
propagation in an effective potential. If the effective potential
has a well, then waves get `trapped' in the well and amplified by
superradiance, thereby triggering an instability. In the case of
massive fields on a four-dimensional Kerr background, the effective
potential indeed has such a well. Consequently, the massive field
grows exponentially and is unstable. It is the presence of a bound
state that simulates the mirror, and so without a bound state we
should never get an instability.

It was found in \cite{super1,super2,super3} that introducing KK
momentum for a massless field in black string (or brane) geometries
can be equivalent to having a mass and so can also trigger
superradiant instabilities. Hence we pay particular attention to KK
momentum in our analysis below.

The strategy here is to look for bound states in the  effective
potential to ascertain whether or not the geometry is superradiantly
unstable. We start by considering the case in which the waves have
no KK momentum, $\lambda= 0$. In this case, at infinity the
potentials go to zero, as indicated by (\ref{lim V}). In
Fig.~\ref{fig:bh} we sketch two typical examples of potentials that
can occur when $\lambda= 0$. The specific parameters that yield
these plots are black holes with $(a_1=32\,,a_2=16\,,M=1.01
(a_1+a_2)^2\,,c_1=5\,,c_5=1.517\,,c_p=5\,,R=1)$, and modes with
$(l=10\,,m_{\phi}=0\,,\lambda=0)$ and $m_{\psi}=10$
[Fig.~\ref{fig:bh}.a] and $m_{\psi}=-10$ [Fig.~\ref{fig:bh}.b]. The
geometry has superradiant modes that satisfy (\ref{super cond}), and
examples of these are represented by a dashed line in these plots.
In Fig.~\ref{fig:bh}.a, an incident wave with frequency $\omega_{\rm
sup}<\omega< 0$ is reflected in the potential $V_-$ and returns back
to infinity with an amplified amplitude. Similarly, in
Fig.~\ref{fig:bh}.b, superradiant modes are those with $0<\omega<
\omega_{\rm sup}$ that are reflected in the potential $V_+$. In both
cases, the pattern speed of the superradiant modes,
$\Sigma_{\psi}=\omega/(R m_{\psi})$, is negative. This means that to
be amplified, the waves must satisfy (\ref{super cond}) and, in
addition, rotate in the same sense as the black hole rotation (the
angular velocity of the black hole $\Omega_{\psi}$ is negative --
see \eqref{velocity}). The second necessary condition for the
superradiant instability -- the existence of bound states -- is
however absent in these plots. We have done an extensive search (by
varying the parameters $a_i\,,M\,,c_i\,,R$ and
$l\,,m_{\psi}\,,m_{\phi}$) for superradiant bound states and have
found none with $\lambda=0$. For the sake of clarity, a typical
example of Schr$\ddot{\rm o}$dinger potentials that would allow
superradiant bound states is sketched in Fig.~\ref{fig:unstable}.a.

\begin{figure}[h]
\begin{center}
\resizebox{14cm}{12cm}
 {\includegraphics{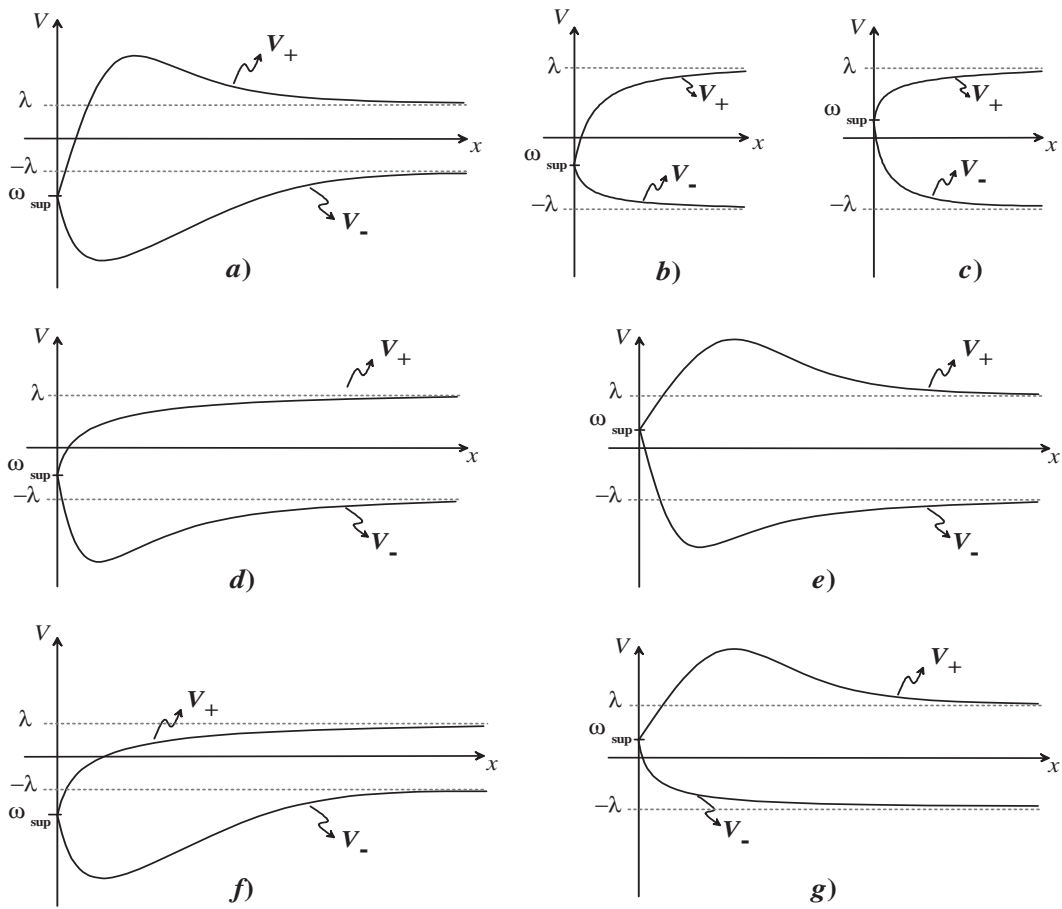}}
\end{center}
\caption{Black hole branch,  $M\geq (a_1+a_2)^2$ or $r_+^2>0$, for
modes with  KK momentum, $\lambda\neq 0$. Although the KK momentum
provides a barrier at infinity, it is not able to create bound
states where superradiant unstable modes could eventually live. }
 \label{fig:bh lambda}
\end{figure}

We now turn our attention to modes with KK momentum, \ie  with
$\lambda\neq 0$. The KK momentum provides a potential barrier of
weight $\lambda$ at infinity [see (\ref{lim V})] and thus are the
most promising when looking for instabilities. However, not even in
this case does there seem to be an instability, as shown in
Fig.~\ref{fig:bh lambda}. This figure represents the several shapes
of potentials $V_{\pm}$ that we can get when $\lambda\neq 0$. None
of these plots has bound states and so it seems that the D1-D5-P
black holes are not afflicted by the superradiant instability. In
Fig.~\ref{fig:unstable}.b, we plot an example of a case for which
the system with $\lambda\neq 0$ would have superradiant
instabilities. As we just stated, we found no such case. In Table
\ref{tab}, we give specific examples of parameters that have the
Schr$\ddot{\rm o}$dinger potentials presented in Fig.~\ref{fig:bh
lambda}.

\begin{table}
\begin{center}
\begin{tabular}{|c|c|c|c|c|c|c|c|c|c|c|c|c|}
\hline
$\bm \lambda $& 0 & $\pm 0.0005$ & $ -0.005$ & $0.005 $  & $ -0.05$  & $0.05 $ & $-0.5 $ &
$ 0.5$ & $-1 $ & $1 $ & $-5 $ & $5 $ \\
\hline
 {\bf Figure} & Fig.~\ref{fig:bh}.a & Fig.~\ref{fig:bh lambda}.a & Fig.~\ref{fig:bh lambda}.e &
 Fig.~\ref{fig:bh lambda}.a
    & Fig.~\ref{fig:bh lambda}.g  & Fig.~\ref{fig:bh lambda}.f & Fig.~\ref{fig:bh lambda}.g
     & Fig.~\ref{fig:bh lambda}.d & Fig.~\ref{fig:bh lambda}.g & Fig.~\ref{fig:bh lambda}.d
       & Fig.~\ref{fig:bh lambda}.c & Fig.~\ref{fig:bh lambda}.b \\
\hline
  \end{tabular}
\end{center}
\caption{\label{tab} Some examples of modes that have the
Schr$\ddot{\rm o}$dinger potentials plotted in Fig.~\ref{fig:bh
lambda}. The black hole geometry is described by the parameters
$(a_1=32\,,a_2=16\,,M=1.01
(a_1+a_2)^2\,,c_1=5\,,c_5=1.517\,,c_p=5\,,R=1)$. The modes have
$(l=m_{\psi}=10\,,m_{\phi}=0)$ and $\lambda$ indicated in the first
row  of the table. Similar Schr$\ddot{\rm o}$dinger potentials are
obtained when $m_{\phi}$ is switched on.}
\end{table}

As discussed in \cite{super2} the absence of bound states in the
potential seems to be closely related to the absence of stable
circular orbits in the geometries, which is a generic feature of
higher dimensional spaces \cite{super1,super2,super3}. It is not
so surprising then to find that higher dimensional rotating black
holes are stable against the superradiant mechanism.

We end this section by considering perturbations of a massive free
scalar. Massive scalar fields, with mass $\mu$, obey the equation
\beq \frac{1}{\sqrt{-g}}\partial_A\left (\sqrt{-g}g^{AB}\partial_B
\Psi\right ) =\mu^2\Psi \,, \label{klein2} \eeq
with the determinant given by Eq. (\ref{gString6}). It is easy to
see that, under the same separation ansatz (\ref{separation
ansatz}), the wave equation will be separable if and only if
$\sqrt{-g}$ itself separates. This happens in particular if ${\tilde
H_1}={\tilde H_5}$, \ie if $Q_1=Q_5$.\footnote{This condition is
also satisfied to a good approximation, if $a_1/r_+$ and $a_2/r_+$
are very small.} In this case, under the assumption of small mass
$\mu$, the angular equation is still given by \eqref{angeq} with the
angular eigenvalue $\Lambda$ defined in \eqref{app}, and the radial
wave equation \eqref{radialeq-r} receives now a source term on its
right-hand side given by $\mu^2(r^2+Ms_1^2)h$. We followed the same
procedure to turn this equation into a Schr\"{o}dinger-like ODE.
Again, we did not find evidence of bound states. Therefore the
inclusion of a scalar mass does not seem sufficient to make these
geometries unstable through the superradiant mechanism.

\section{\label{sec:Conc}Discussion}

In this paper, we studied a general family of supergravity solutions
for the D1-D5-P system which contains two special branches: one with
smooth horizon-free geometries (\jmart solitons) and a black hole
branch. In general, when rotating, these solutions have an
ergoregion. This is absent only in the supersymmetric case (in both
branches). However, we note that if these backgrounds are regarded
as solutions of type I supergravity, some of the ergo-free
backgrounds are non-supersymmetric (on both branches), \eg certain
solutions are extreme but non-supersymmetric black holes
 \cite{Dias:2007dj,DES2007}. With an ergoregion, the background
exhibits superradiant phenomena, however, superradiant scattering
{\it per se} is harmless. It will only extract a small amount of
rotational energy from the geometry and transfer it to a wave that
transports it to infinity. There are, however, two situations where
the addition of extra ingredients leads to a catastrophic
instability. One is when we have a spacetime geometry with an
ergoregion but no horizon. This situation generically leads to the
so-called ergoregion instability \cite{friedman} in which there are
modes which are outgoing at infinity, regular at the origin, and
growing unboundedly with time. The negative energy that is stored in
the ergoregion core, by energy conservation, will then also grow
negative without bound. In a previous article, we found that the
D1-D5-P smooth horizon-free geometries are afflicted by this
instability \cite{Cardoso:2005gj}.

The other scenario which produces a catastrophic instability is more
well-known. A rotating black hole with an ergosphere and with a
`reflecting wall' can lead to multiple superradiant
scattering/reflection that extracts rotational energy from the black
hole without bound. This reflecting wall can be provided by an
artificial mirror \cite{press}, by the mass of the propagating field
\cite{detweiler}, by an asymptotic anti-de Sitter geometry
\cite{CardDiasAdS} or by the KK momentum of a black string geometry
\cite{super1,super2,super3}. The key feature, necessary for the
activation of this instability, is that the effective potential that
describes the field propagation in the given background must have a
well where bound states can be trapped. The D1-D5-P black hole could
potentially be afflicted by this superradiant instability. Indeed,
it has both an ergoregion and KK momentum along the distinguished
$S^1$ circle, which might create the mentioned reflecting boundary.
However, in our extensive search over the parameters of the
solution, we have found that the other ingredient -- the potential
well -- is absent. Hence the D1-D5-P black holes do not appear to
suffer from any superradiant instability.

In sections \ref{sec:WaveSep} and \ref{sec:Sch}, we verified the
absence of a supperradiant instability for a minimally coupled
scalar field (both massless and massive) propagating in the D1-D5-P
black hole background. Of course, it would be most interesting to
verify that the black hole remains stable when we perturb it by the
fields of the type IIb supergravity theory. Hence, we now address
the question of to what extent our analysis applies to the
supergravity fields. Of course, the same discussion applies to our
instability analysis of the \jmart solitons \cite{Cardoso:2005gj}
which was also explicitly carried out with a massless minimally
coupled field.

As discussed at the beginning of section \ref{sec:WaveSep}, we are
considering the propagation of scalars in the six-dimensional space
comprised of the non-compact directions and the distinguished $S^1$
circle. Hence let us consider the scalars appearing in the reduced
six-dimensional supergravity. After compactifying the type IIb
supergravity down to six dimensions, the scalars parameterize the
moduli space
\beq {\cal M}_0= SO(5,n;R)/SO(5)\times SO(n) \label{mod0}\eeq
up to additional global identifications, where $n$ = 5 or 21 for
$M^4=T^4$ or $K3$, respectively \cite{whitehouse}. Hence the
six-dimensional theory contains a total of $5n$ independent scalars.
However, when the D1-D5 string is introduced in the six dimensions,
several of these scalars interact with the RR two-form \reef{rr}
sourced by the string. These `fixed' scalars acquire an effective
mass in the $AdS_3\times S^3$ core of the black hole. Setting aside
these fixed scalars, the residual moduli space is
\beq {\cal M}= SO(4,n;R)/SO(4)\times SO(n) \label{mod1}\eeq
again up to additional global identifications \cite{whitehouse}.
Hence we are left with $4n$ `minimal' scalars in the six-dimensional
D1-D5-P black string background.

In the six-dimensional effective action, the kinetic term for the
scalars can be written as \cite{skalar}:
\beq I_{scalar}= \int d^{6} x \sqrt{{-g}} \,g^{\mu\nu} L_{ij}
\partial_\mu M^{jk} L_{kl} \partial_\nu M^{li}\,,
\label{scalact}\eeq
where the scalar fields are represented by the $(5+n)\times(5+n)$
matrix $M^{ij}$ taking values in the coset \reef{mod0}. These then
satisfy the following identities:
\beq M^T=M\qquad{\rm and} \qquad M\,L\,M^T=L\qquad {\rm with}\
L_{ij}=~
 \begin{pmatrix}
 ~0~&{\bf 1_{\ssc 5}} &~0~ \cr
 {\bf 1_{\ssc 5}}&~0~&~0~\cr
 ~0~&~0~& {\bf 1_{\ssc n-5}}
 \end{pmatrix}
\label{ldef} \eeq
where ${\bf 1}_{\ssc d}$ are $~ d \times d~$ unit matrices and the
superscript $T$ indicates matrix transposition. As discussed in
section \ref{sec:WaveSep}, the six-dimensional dilaton vanishes and
hence in eq.~\reef{scalact}, $g_{\mu\nu}$ corresponds to both the
string-frame or Einstein-frame metric. The full scalar action also
includes couplings to the gauge fields and form fields in six
dimensions \cite{skalar}, as well as to internal fluxes
\cite{odd,oddrev}. As discussed above, in the present background,
these interactions are irrelevant for the $4n$ minimal scalars and
so eq.~\reef{scalact} compromises the entire effective action for
these fields. Now this action \reef{scalact} has a deceptively
simple form and so one might conclude that all of the minimal
scalars satisfy the massless Klein-Gordon equation \reef{klein}.
However, in fact, the action is implicitly nonlinear since the
scalars take values on the coset \reef{mod1} and so in general, this
conclusion is mistaken. For example, with the torus
compactification, \ie $M^4=T^4$, the equation of motion for scalars
originating from the internal components of the ten-dimensional
metric is
\beq \nabla^2 G^{ab}-G^{ac}\,G^{bd}\,\nabla^2 G_{cd}=0\,.
\label{intmet} \eeq
when the other fields are set to zero. Carefully examining an
explicit representation of $M$ \cite{skalar}, one finds that
relatively few of the scalars actually satisfy eq.~\reef{klein} in
general. Recall that we must focus on the minimal scalars of
\reef{mod1} and then there are just three scalars comprising the
antisymmetric tensor parameterizing the 4 $\times$ 4 block just
above the diagonal in the top-right corner of $M$ \cite{skalar}, \ie
in the same position as ${\bf 1}_{\ssc 5}$ appears in $L$. In the
torus compactification, these three scalars correspond to the
internal components of the RR two-form, $C^{(2)}_{ab}$ and beginning
with the $d=10$ supergravity action, one can easily demonstrate that
these field obey the ordinary massless Klein-Gordon equation in six
dimensions. The general discussion here confirms that in fact these
three fields are the only scalars satisfying this simple equation of
motion, in general. Of course, there is one case which deserves
special attention: $Q_1=Q_5$. In this case, the background scalars,
\ie the ten-dimensional dilaton \reef{dilatone} and the internal
volume \reef{eta}, are constant. This trivializes the nonlinearities
in the scalar field equations and any linear perturbations of the
minimal scalars about the background satisfy the massless
Klein-Gordon equation \reef{klein}. For example, for the $T^4$
compactification, the internal moduli may be written as
$G_{ab}=\delta_{ab}+h_{ab}$ with traceless perturbations $h_{ab}$
and the eq.~\reef{intmet} reduces to $\nabla^2h_{ab}=0$.

The lesson we derive from the above discussion is that there are
precisely three fields in the six-dimensional supergravity whose
fluctuations are described by the Klein-Gordon equation
\eqref{klein} analyzed in Sections \ref{sec:WaveSep} and
\ref{sec:Sch}, as well as in \cite{Cardoso:2005gj}. However, in the
special case $Q_1=Q_5$, all $4n$ minimal scalars satisfy this
equation. Therefore, our conclusions that we derived for the
minimally coupled (massless) scalar field apply straightforwardly
for these fields above. Further then, and as the main conclusion,
perturbations of certain supergravity fields can drive the \jmart
geometries \cite{ross} unstable due to the ergoregion instability
\cite{Cardoso:2005gj} but on the other hand, these same fields do
not seem to produce a comparable (superradiant) instability for the
D1-D5-P black holes.

While our analysis does not apply in general to the remaining $5n-3$
scalars, we are tempted to discuss these in qualitative terms. Quite
generally, we expect that the superradiant instability will not
appear for these scalars, independent of most of the details of
their wave equation. First, we observe that for very short
wavelengths, we expect wave packets to propagate along the
characteristic curves of the wave equation. For example, with the
Klein-Gordon equation \eqref{klein}, high-frequency wave packets
travel along null geodesics. These characteristics are determined by
the principle part of the wave equation, \ie the second order terms.
Now for a general scalar, we expect that these characteristics will
in fact match the null geodesics of the Klein-Gordon field. The
relevant term in the action will be precisely the kinetic term of a
given scalar \eqref{scalact}. Hence we can note here that the
interactions with the background RR two-form will not affect the
characteristics. Next the background scalars will only modify the
kinetic term of any given scalar excitation by multiplying the
latter with a nontrivial overall factor. Hence the principle part of
the resulting wave equation is only modified by an overall factor
which leaves the characteristics unchanged. Hence we expect that
short-wavelength wave packets of all of $5n$ scalars propagate along
null geodesics of the six-dimensional geometry.

Next we consider the analysis of \cite{super1,super2} which
considered the superradiant instability in black strings and black
branes in arbitrary dimensions -- these were solutions of the vacuum
Einstein equations. It was found that while an instability appears
with four (noncompact) dimensions, no such instability appears in
five and higher dimensions. Only in four dimensions did the
background provide an effective potential which trapped bound
states. However, it was pointed that in the high-frequency limit
such a bound state would be following a stable circular orbit in the
black hole background and so in higher dimensions, the absence of
bound states can be related to the absence of stable circular
orbits, as mentioned in section \ref{sec:Sch}. In the present case,
we again are studying black strings in six dimensions or effectively
five-dimensional black holes. Hence we should not expect to find
stable orbits in these backgrounds. While a complete proof would
require a new detailed analysis, the absence of a trapping potential
for Klein-Gordon scalars certainly suggests the absence of any such
orbits. The absence of such orbits can then be used to argue the
absence of bound states and hence the absence of a superradiant
instability for a general supergravity scalar.

In passing we note that the existence of negative-energy geodesics
trapped in the ergoregion of the \jmart solutions can be argued on
general grounds, as discussed in \cite{Cardoso:2005gj}. Hence a
similar reasoning to that above suggests that any of the
supergravity scalars can initiate the ergoregion instability in
these backgrounds. That is, these bound geodesics would correspond
to trapped states in the context of a field theory analysis. The key
question then becomes whether the corresponding scalar field modes
of the field 'fit' inside the ergoregion or whether they 'leak' out
to infinity, i.e., whether they correspond to a true negative-energy
bound state or to a mode producing the ergoregion instability. The
detailed analysis of \cite{Cardoso:2005gj} showed that both kinds of
modes existed for a Klein-Gordon scalar but in particular the
spinning modes were generically associated with the ergoregion
instability. While we have not extended this detailed analysis to
the complete collection of six-dimensional supergravity scalars, we
expect that the similar results would be found. That is, the
ergoregion instability will generically be initiated by such modes
that are `trapped' by their angular momentum.

Of course, the primary motivation for the present study were the
possible implications for Mathur's fuzzball proposal \cite{fuzzy}.
We have already presented an extensive discussion on this topic in
\cite{Cardoso:2005gj} and will only comment on a few of the salient
points here. According to Mathur's proposal, the individual
microstates of a black hole are described by                   
 horizon-free solitons and the black hole geometry only appears after `averaging'
over these microstate solutions. Much of the evidence for these
ideas comes from studying certain supersymmetric solutions in five
\cite{two,three12,three2} and four \cite{four} dimensions. However,
if the fuzzball proposal is to have any substance, it must also
extend to non-supersymmetric black holes. The \jmart solitons
provide the first family of smooth horizon-free geometries which are
non-supersymmetric and so correspond to non-BPS microstates
\cite{ross}. However, as mentioned in the introduction, these
solutions also present an apparent contradiction with the fuzzball
proposal. That is, the \jmart solitons suffer from a classical
instability, namely an ergoregion instability, and further it can be
argued that this instability should be robust feature of any smooth
horizon-free geometry carrying angular momentum
\cite{Cardoso:2005gj}. However, the nonextremal rotating D1-D5-P
black holes exhibit no comparable instability. In particular, we
have shown here that these black holes do not exhibit a superradiant
instability. Hence there is a possible contradiction for the
fuzzball proposal since one would expect that if the ergoregion
instability is common to all of the rotating non-BPS microstate
geometries, then this instability should be reflected in the black
hole geometry which is supposed to arise from averaging over the
microstate solutions.

However, this reasoning is not definitive and this puzzle still has
physically sensible resolution, at least in principle. In
particular, there are two observations which we must make about the
\jmart solitons. First, the mass and spin of the \jmart and black
hole solutions are in very different regimes, as described in
\eqref{branches}: $M \leq(a_1-a_2)^2$ for the \jmart branch and
$M\geq (a_1+a_2)^2$ for the black hole branch. Hence the \jmart
solutions should be expected to represent at best a very small
contribution to the microstate ensemble underlying a nonextremal
D1-D5-P black hole. Second, the \jmart solutions are very symmetric
spacetimes. In particular, they have all of the same Killing
symmetries as the D1-D5-P black holes, since these are simply two
branches of a common family of supergravity solutions. In contrast,
generic microstate geometries are expected to contain complex
throats which do not respect these Killing symmetries \cite{fuzzy}.
Hence the physical characteristics of the \jmart solutions are
likely not representative of those for a typical microstate in the
black hole ensemble. While the typical microstate geometries should
still suffer from an ergoregion instability, one might expect that
the instability timescale becomes very long \cite{Cardoso:2005gj},
especially in the `classical limit' where the string coupling is
taken to zero \cite{class}. In particular, the complex throat at the
core of the typical microstate geometries should emulate the
absorptive behaviour of the black horizon in this limit, making it
difficult to distinguish physics in these backgrounds from that in a
black hole background, except on very large time scales. It is
reasonable then that the timescale of the ergoregion instability
should be a scale that grows parametrically as $g_s\rightarrow0$.

Given that the ergoregion instability should be a generic feature of
nonsupersymmetric microstate geometries, it would be interesting to
study the dual non-BPS microstates at weak coupling for evidence of
such an instability. As a particularly simple set of microstates
have been identified to correspond to the \jmart solitons
\cite{ross}, these may provide a good framework to initiate such a
line of investigation. For more general microstates, \eg those
expected to describe a near-extremal spinning D1-D5-P black hole
\cite{near}, one must be careful to distinguish the expected Hawking
radiation from radiation related to the ergoregion instability. In
this regime, the latter is likely to be related to the `nonthermal'
radiation that is expected to produce the spin-down of the black
hole \cite{page}. It may be, however, that when considering non-BPS
configurations that the ergoregion instability provides a signature
by which microstate geometries are more easily distinguished from
their black hole counter-parts. Hence this seems a promising
direction of research.

Of course, another challenging problem which remains is the
construction of a more or less complete family of microstate
geometries beyond the BPS sector. While the existence of the \jmart
solitons indicates that at least certain non-BPS states can be
described by                                
 horizon-free geometries, it is not at all clear
that this property should be shared by all such states. However,
this is certainly another intriguing research direction.

\section*{Acknowledgements}
It is a pleasure to thank Gary Horowitz, Vishnu Jejjala and Simon
Ross for interesting discussions. We also thank Jordan Hovdebo for
collaborating with us at an early stage of this project. OJCD would
like to thank both the Kavli Institute for Theoretical Physics and
the Perimeter Institute for Theoretical Physics for their
hospitality during various stages of this research. This work was
partially funded by Fundac\~ao para a Ci\^encia e Tecnologia through
project PTDC/FIS/ 64175/2006. VC acknowledges financial support from
the Funda\c c\~ao Calouste Gulbenkian through Programa Gulbenkian de
Est\'{\i}mulo \`a Investiga\c c\~ao Cient\'{\i}fica. OJCD
acknowledges financial support provided by the European Community
through the Intra-European Marie Curie contract MEIF-CT-2006-038924.
Research at the Perimeter Institute is supported in part by the
Government of Canada through NSERC and by the Province of Ontario
through MRI. RCM acknowledges support from an NSERC Discovery grant
and from the Canadian Institute for Advanced Research. Research at
the KITP was supported in part by the NSF under Grant No.
PHY05-51164. 

\appendix
\section{\label{sec:JMaRT}Simplification of the wave equation for the JM\lowercase{a}RT soliton}


In this appendix we clarify the connection between the present
perturbation analysis on the black hole branch of the D1-D5-P system
and the stability analysis of the \jmart solitons done in
\cite{Cardoso:2005gj}.

In either case, we require that the function $g(r)$, given in
eq.~\reef{def function g}, has real roots. The \jmart solitons then
appear in the low-mass regime of (\ref{branches}), $M^2\leq
(a_1-a_2)^2$, where $r_+^2<0$. While the metric may appear singular
at $r^2=r_+^2$, one can impose a series of constraints that ensure
that the solutions are free of singularities, horizons and closed
time-like curves. This task leads to the construction of the \jmart
solitons \cite{ross}. These solitonic solutions have an appropriate
circle that shrinks to zero at the origin and the constraints ensure
that this happens smoothly. First, $M$ and $R$ are re-expressed in
terms of the remaining parameters -- see Eqs.~(3.15) and (3.20) of
\cite{ross},
\begin{eqnarray}
& & R =\frac{M s_1 c_1 s_5 c_5 \sqrt{s_1 c_1 s_5 c_5 s_p c_p}}
{\sqrt{a_1 a_2} (c_1^2 c_5^2 c_p^2-s_1^2 s_5^2 s_p^2)}\,,
\nonumber \\
& &  M= a_1^2+a_2^2-a_1 a_2\,
 \frac{c_1^2 c_5^2 c_p^2+s_1^2 s_5^2 s_p^2}{s_1 c_1 s_5 c_5 s_p
 c_p}\,.
 \label{JMaRTconstraints}
\end{eqnarray}
Then ensuring the geometry remains smooth requires imposing to
`quantization' conditions
\begin{eqnarray}
\frac{s_p c_p}{a_1 c_1 c_5 c_p- a_2 s_1 s_5 s_p}\,R&=&n \nonumber \\
-\frac{s_p c_p}{a_2 c_1 c_5 c_p- a_1 s_1 s_5 s_p}\,R&=&m
 \label{JMaRTconstraints2}
\end{eqnarray}
where $m,n$ are both integers \cite{ross}. These two constraints can
be put in a more elegant form by introducing the dimensionless
quantities,
\beq j=\left ( \frac{a_2}{a_1}\right )^{1/2} \leq 1 \,,\qquad s
=\left ( \frac{s_1 s_5 s_p}{c_1 c_5 c_p} \right )^{1/2}
\,,\label{dimless}\eeq
with which the constraints \reef{JMaRTconstraints} can be
re-expressed as
\beq \frac{j+j^{-1}}{s+s^{-1}}=m-n\,, \qquad
\frac{j-j^{-1}}{s-s^{-1}}=m+n\,.
 \label{JMaRTconstraints1}\eeq
Again, without loss of generality, we have assumed $a_1 \ge a_2\ge
0$, which further implies $m>n\ge0$. We also note here that the
special case $m=n+1$ corresponds to supersymmetric solutions. In
this case one also has: $M=0$, $s=1$, $j=1$, $a_1=a_2$.

Imposing the constraints \reef{JMaRTconstraints2} leaves a
five-parameter family of smooth solitonic solutions. We can think of
the independent parameters as the D1-brane and D5-brane charges,
$Q_1,Q_5$; the (asymptotic) radius of the $y$-circle, $R$; and the
two integers, $m$ and $n$, which fix the remaining physical
parameters as \cite{ross}
\begin{equation}
Q_P=nm\frac{Q_1Q_5}{R^2}\,,\quad J_\phi=-m\frac{Q_1Q_5}{R}\,,\quad
J_\psi=n\frac{Q_1Q_5}{R}\,. \label{simple}
\end{equation}
Of course, depending on the specific application, it may be more
appropriate and/or simpler to describe the solutions using a
different set of quantities. To conclude our discussion of the
\jmart case, we note that the roots (\ref{r+-}) of $g^{rr}$ can be
rewritten as
\begin{eqnarray}
r_+^2=- a_1 a_2 \frac{s_1 s_5 s_p}{c_1 c_5 c_p}\,, \qquad  r_-^2=-
a_1 a_2 \frac{c_1 c_5 c_p}{s_1 s_5 s_p}\,,
 \label{JMaRTconstraints3}
\end{eqnarray}

The wave equation in the background of the \jmart solitons is still
given by (\ref{radialeq-r}), but we can simplify it by using the
\jmart constraints
(\ref{JMaRTconstraints})-(\ref{JMaRTconstraints3}). The results of
this Appendix will make use of these constraints and so they will be
valid only for the horizon-free \jmart solutions; they no longer
apply to the general case and in particular to the black hole case.
Define
\begin{eqnarray}
& & \rho=\frac{c_1^2 c_5^2 c_p^2-s_1^2 s_5^2 s_p^2}{s_1 c_1 s_5
c_5} \,,\nonumber \\
& & \vartheta =\frac{c_1^2 c_5^2 -s_1^2 s_5^2}{s_1 c_1 s_5
c_5}\,s_p c_p\,,
 \label{JMaRTconstraints4}
\end{eqnarray}
and note that  $r_+^2-r_-^2=\frac{a_1 a_2 \rho}{s_p c_p} $. If and
only if the \jmart constraints are imposed, we can verify the
following identities:
\begin{eqnarray}
\begin{aligned} &
 -\frac{\omega^2}{g(r)} \,\frac{M^2}{R^2} \left[ - c_1^2 c_5^2
c_p^2  r^2
  - 2 s_1 s_5 s_p c_1 c_5 c_p  a_1 a_2
  + s_1^2 s_5^2 s_p^2  (r^2+a_1^2+a_2^2-M)\right]
   = \frac{\omega^2}{g(r)} \, (r_+^2-r_-^2) \rho^2 (r^2-r_+^2)  \,,
  \\
  & \frac{2\omega\lambda}{g(r)} \,\frac{M^2}{R^2}\,\left[  -s_p c_p \left[-c_1^2
c_5^2   r^2 + s_1^2 s_5^2  (r^2+a_1^2+a_2^2-M)\right]+
(c_p^2+s_p^2) s_1 s_5 c_1 c_5  a_1 a_2\right]
 =  \frac{2\omega\lambda}{g(r)} \, (r_+^2-r_-^2) \rho \vartheta (r^2-r_+^2)\,,
 \\
 & -\frac{2\omega m_{\phi}}{g(r)}\, \frac{M}{R}  \left[ -c_1
c_5 c_p  a_2 (r^2+a_1^2) + s_1 s_5 s_p  a_1 (r^2+a_1^2-M) \right]
 = -\frac{2\omega m_{\phi}}{g(r)}\, (r_+^2-r_-^2) \rho n (r^2-r_+^2) \,,
 \\
 & \frac{2\omega m_{\psi}}{g(r)}\, \frac{M}{R} \left[ c_1
c_5 c_p a_1 (r^2+a_2^2) - s_1 s_5 s_p  a_2 (r^2+a_2^2-M) \right]
 = \frac{2\omega m_{\psi}}{g(r)}\, (r_+^2-r_-^2) \rho m (r^2-r_+^2)\,,
 \\
 & - \frac{\lambda^2}{g(r)} \,\frac{M^2}{R^2} \left[ - c_1^2 c_5^2 s_p^2  r^2
  - 2 s_1 s_5 s_p c_1 c_5 c_p  a_1 a_2
  + s_1^2 s_5^2 c_p^2  (r^2+a_1^2+a_2^2-M) \right] \\
 & \hspace{6cm} = -\frac{\lambda^2}{g(r)} \,(r_+^2-r_-^2) \left[(r^2-r_-^2) - \vartheta^2 (r^2-r_+^2) \right] \,,
\\
 &   \frac{2\lambda m_{\phi}}{g(r)}\, \frac{M}{R} \left[ c_1
c_5 s_p  a_2 (r^2+a_1^2) - s_1 s_5 c_p  a_1 (r^2+a_1^2-M) \right]
  = \frac{2\lambda m_{\phi}}{g(r)}\, (r_+^2-r_-^2) \left[-m(r^2-r_-^2) - n \vartheta (r^2-r_+^2) \right]  \,,\\
 & \frac{2\lambda m_{\psi}}{g(r)}\, \frac{M}{R}\, \left[ c_1
c_5 s_p  a_1 (r^2+a_2^2) - s_1 s_5 c_p  a_2 (r^2+a_2^2-M) \right]
  = \frac{2\lambda m_{\psi}}{g(r)}\, (r_+^2-r_-^2) \left[n(r^2-r_-^2) + m \vartheta (r^2-r_+^2) \right] \,,\\
 & - \frac{m_{\phi}^2}{g(r)}\, \left[ (r^2+a_1^2)(a_1^2-a_2^2)-M a_1^2\right]
  = - \frac{m_{\phi}^2}{g(r)}\, (r_+^2-r_-^2) \left[m^2(r^2-r_-^2) - n^2  (r^2-r_+^2) \right]  \,,\\
 & \frac{2m_{\phi}m_{\psi}}{g(r)}\, M a_1 a_2
  = \frac{2m_{\phi}m_{\psi}}{g(r)}\,(r_+^2-r_-^2)^2 n m \,,\\
 & - \frac{m_{\psi}^2}{g(r)}\,  \left[(r^2+a_2^2)(a_2^2-a_1^2)-M a_2^2\right]
  = - \frac{m_{\psi}^2}{g(r)}\, (r_+^2-r_-^2) \left[n^2(r^2-r_-^2) - m^2  (r^2-r_+^2) \right] \,.
 \label{VerifyWaveEqJMaRT}
 \end{aligned}
\end{eqnarray}
So, in these equalities, the left-hand side was taken from
(\ref{radialeq-r}) and the right-hand side is valid only if the
JMaRT constraints (\ref{JMaRTconstraints})-(\ref{JMaRTconstraints2})
are imposed. Hence in the \jmart case, we leave the first line of
(\ref{radialeq-r}) unchanged but we can simplify all the terms
proportional to $g(r)$. Inserting (\ref{VerifyWaveEqJMaRT}) into
eq.~(\ref{radialeq-r}) yields the simplified version of the wave
equation for the \jmart solitons,
\begin{eqnarray}
\begin{aligned} & \frac{1}{r} \frac{d}{dr} \left[ \frac{g(r)}{r}
\frac{d}{dr} h \right] - \Lambda h + \left[ \frac{(\omega^2 - \lambda^2)}{R^2} (r^2 + M s_1^2 + M s_5^2) +
(\omega c_p + \lambda s_p)^2 \frac{M}{R^2} \right] h
\\
& -(r_+^2-r_-^2)\frac{(\lambda - n m_\psi + m m_\phi)^2}{(r^2-r_+^2)} \, h + (r_+^2-r_-^2)\frac{(\omega \varrho
+ \lambda \vartheta - n m_\phi + m m_\psi)^2}{(r^2- r_-^2)} \, h = 0 \,, \label{WaveEqJMaRT}
\end{aligned}
\end{eqnarray}
which is exactly Equation (6.4) of \cite{ross} and Equation (14)
of \cite{Cardoso:2005gj}.

\section{\label{sec:potential}The Schr\"{o}dinger potentials}

In this appendix we give the explicit form of the function of the
function $\gamma$ and of the Schr$\bm \ddot{\rm o}$dinger
potentials $V_{\pm}$ that are defined in (\ref{Vfactorized}):
\begin{eqnarray}
 \gamma&=&\frac{M^2}{r^6R^2}\left[ c_1^2c_5^2c_p^2\,r^2+\frac{g(r)}{M^2}\left (r^2+M(c_p^2+s_1^2+s_5^2)\right )+
2a_1a_2c_1c_5c_ps_1s_5s_p+\left (M-a_1^2-a_2^2-r^2\right )s_1^2s_5^2s_p^2\right ]\,,\nonumber \\
V_{\pm}&=&-\frac{K_1}{2\gamma} \pm \sqrt{ \left (
\frac{K_1}{2\gamma} \right )^2 -\frac{K_0}{\gamma} }\,, \label{V+V-}
 \end{eqnarray}
with
\begin{eqnarray}
K_0 &=&\frac{g(r)}{4r^{10}}\left [-3g(r)-12 \left (-2r_-^2
r_+^2+r^2(r_-^2+r_+^2)\right )+4r^4
\left (l(l-2)-\frac{\lambda^2}{R^2}(-Ms_p^2+r^2+Ms_1^2+Ms_5^2)\right )\right ] \nonumber \\
& & + \frac{M^2\lambda^2}{R^2r^6}\left [
    c_1^2c_5^2s_p^2r^2+2a_1a_2c_1c_5c_ps_1s_5s_p-c_p^2s_1^2s_5^2
    (a_1^2a_2^2-M+r^2)\right ]\nonumber \\
& & +\frac{2M\lambda}{Rr^6} \left [-c_ps_1s_5m_{\phi}\left (
    a_1^3+a_1(r^2-M)+a_2\frac{m_{\psi}}{m_{\phi}}(r^2-M+a_2^2)\right)
    +a_1a_2c_1c_5s_pm_{\phi}\left(
    a_1+a_2\frac{m_{\psi}}{m_{\phi}}+\frac{r^2}{a_1}+\frac{r^2m_{\psi}}
    {a_2m_{\phi}}\right) \right ]\nonumber \\
 & &-\frac{1}{r^6}\left[
r^2(a_1^2-a_2^2)(m_{\phi}^2-m_{\psi}^2)+(a_1^2m_{\phi}^2-a_2^2m_{\psi}^2)
\left ( a_1^2-a_2^2-M\right ) \right] \,,
                                        \nonumber \\
K_1 &=&\frac{2M}{r^6R}{\biggl [} c_1c_5c_p\left[
a_1a_2(a_1m_{\phi}+a_2m_{\psi})+r^2(a_2m_{\phi}+a_1m_{\psi})\right]
\nonumber \\
& & \hspace{1cm} -\left [a_1^3m_{\phi}+a_1m_{\phi}(r^2-M)
      +a_2m_{\psi}(a_2^2-M+r^2)\right ]s_1s_5s_p {\biggl ]}\nonumber\\
& & +\frac{2M \lambda}{r^6R^2}{\biggl [}
a_1a_2c_1c_5c_p^2s_1s_5M+c_p s_p\left [c_1^2c_5^2Mr^2 +g(r)-M
s_1^2s_5^2 \left(a_1^2+a_2^2-M+r^2 \right)  \right ]
+a_1a_2c_1c_5Ms_1s_5s_p^2  {\biggl ]} \,. \label{V0V1}
\end{eqnarray}



\begin{thebibliography}{99}

\bibitem{fuzzy} For a review see S.D.~Mathur, ``The fuzzball proposal for black holes:
An elementary review,'' Fortsch.\ Phys.\  {\bf 53}, 793 (2005),
hep-th/0502050;\\
S.D.~Mathur, ``The quantum structure of black holes,''
hep-th/0510180.

\bibitem{two} V.~Balasubramanian, J.~de Boer, E.~Keski-Vakkuri, and S.F.~Ross,
``Supersymmetric conical defects: Towards a string theoretic
description of black hole formation," Phys. Rev. D {\bf 64} (2001)
064011 [arXiv:hep-th/0011217];\\
J.M.~Maldacena and L.~Maoz, ``De-singularization by rotation," JHEP
{\bf 12} (2002) 055 [arXiv:hep-th/0012025];\\
O.~Lunin and S.D.~Mathur, ``Metric of the multiply wound rotating
string," Nucl. Phys. B {\bf 610}  (2001) 49 [arXiv:hep-th/0105136];
  ``AdS/CFT duality and the black hole information paradox,''
  Nucl.\ Phys.\  B {\bf 623} (2002) 342
  [arXiv:hep-th/0109154];
``Statistical interpretation of Bekenstein entropy for systems with
a stretched horizon," Phys. Rev. Lett. {\bf 88} (2002) 211303
[arXiv:hep-th/0202072];
``The slowly rotating near extremal D1-D5 system as a `hot tube',"
Nucl. Phys. B {\bf 615} (2001) 285  [arXiv:hep-th/0107113];
O.~Lunin, J.M.~Maldacena, and L.~Maoz, ``Gravity solutions for the
D1-D5 system with angular momentum," hep-th/0212210;\\
H.~Lin, O.~Lunin and J.M.~Maldacena, ``Bubbling AdS space and 1/2
BPS geometries,''
  JHEP {\bf 0410} (2004) 025
  [arXiv:hep-th/0409174];\\
M.~Taylor, ``General 2 charge geometries,'' hep-th/0507223;\\
K.~Skenderis and M.~Taylor, ``Fuzzball solutions and D1-D5
microstates,''
  Phys.\ Rev.\ Lett.\  {\bf 98} (2007) 071601
  [arXiv:hep-th/0609154];\\
I.~Kanitscheider, K.~Skenderis and M.~Taylor, ``Holographic anatomy
of fuzzballs,''
  JHEP {\bf 0704} (2007) 023
  [arXiv:hep-th/0611171];
``Fuzzballs with internal excitations,''
  arXiv:0704.0690 [hep-th].

\bibitem{three12}  O.~Lunin,
``Adding momentum to D1-D5 system,'' JHEP {\bf 0404} (2004) 054
[arXiv:hep-th/0404006];\\
S.~Giusto, S.D.~Mathur, and A.~Saxena, ``Dual geometries for a set
of 3-charge microstates," Nucl. Phys. B {\bf 701} (2004) 357
[arXiv:hep-th/0405017].

\bibitem{three2} S.D.~Mathur, A.~Saxena and Y.K.~Srivastava,
``Constructing 'hair' for the three charge hole,''
  Nucl.\ Phys.\  B {\bf 680} (2004) 415
  [arXiv:hep-th/0311092];\\
S.~Giusto, S.D.~Mathur, and A.~Saxena, ``3-charge geometries and
their CFT duals," Nucl. Phys. B {\bf 710}  (2005) 425 [arXiv:hep-th/0406103];\\
S.~Giusto and S.D.~Mathur, ``Geometry of D1-D5-P bound states,''
  Nucl.\ Phys.\  B {\bf 729} (2005) 203
  [arXiv:hep-th/0409067];\\
J.~Ford, S.~Giusto and A.~Saxena, ``A class of BPS time-dependent
3-charge microstates from spectral flow,''
  arXiv:hep-th/0612227;\\
I.~Bena and N.P.~Warner, ``One ring to rule them all ... and in the
darkness bind them?,'' hep-th/0408106;
``Bubbling supertubes and foaming black
holes,'' hep-th/0505166;\\
I.~Bena, C.W.~Wang and N.P.~Warner,
  ``The foaming three-charge black hole,''
  Phys.\ Rev.\  D {\bf 75} (2007) 124026
  [arXiv:hep-th/0604110];
``Mergers and typical black hole microstates,''
  JHEP {\bf 0611} (2006) 042
  [arXiv:hep-th/0608217];
``Plumbing the Abyss: Black Ring Microstates,''
  arXiv:0706.3786 [hep-th];\\
P.~Berglund, E.G.~Gimon and T.S.~Levi, ``Supergravity microstates
for BPS black holes and black rings,'' hep-th/0505167.


\bibitem{four} I.~Bena and P.~Kraus, ``Microstates of the D1-D5-KK system,''
Phys.\ Rev.\ D {\bf 72} (2005) 025007 [arXiv:hep-th/0503053];
``Microscopic description of black rings in AdS/CFT,'' JHEP {\bf
0412} (2004) 070 [arXiv:hep-th/0408186].
I.~Bena, P.~Kraus and N.P.~Warner, ``Black rings in Taub-NUT,''
Phys.\ Rev.\ D {\bf 72} (2005) 084019 [arXiv:hep-th/0504142];\\
H.~Elvang, R.~Emparan, D.~Mateos and H.S.~Reall, ``Supersymmetric 4D
rotating black holes from 5D black rings,'' JHEP {\bf 0508} (2005)
042 [arXiv:hep-th/0504125];\\
A.~Saxena, G.~Potvin, S.~Giusto and A.W.~Peet, ``Smooth geometries
with four charges in four dimensions,''
  JHEP {\bf 0604} (2006) 010
  [arXiv:hep-th/0509214];\\
V.~Balasubramanian, E.G.~Gimon and T.S.~Levi, ``Four Dimensional
Black Hole Microstates: From D-branes to Spacetime Foam,''
arXiv:hep-th/0606118.

\bibitem{ross} V.~Jejjala, O.~Madden, S.F.~Ross and G.~Titchener,
``Non-supersymmetric smooth geometries and D1-D5-P bound states,"
Phys. Rev. D {\bf 71}, 1240030 (2005),  hep-th/0504181.

\bibitem{newross} E.G.~Gimon, T.S.~Levi and S.F.~Ross,
``Geometry of non-supersymmetric three-charge bound states,''
arXiv:0705.1238 [hep-th].

\bibitem{Cardoso:2005gj}
  V.~Cardoso, O.J.C.~Dias, J.L.~Hovdebo and R.C.~Myers,
  ``Instability of non-supersymmetric smooth geometries,''
  Phys.\ Rev.\ D {\bf 73}, 064031 (2006); hep-th/0512277.

\bibitem{friedman}
J.L.~Friedman, ``Ergosphere instability," Commun. Math. Phys. {\bf
63}, 243 (1978).

\bibitem{super1}
 V.~Cardoso and J.P.S.~Lemos,
  ``New instability for rotating black branes and strings,''
  Phys.\ Lett.\ B {\bf 621}, 219 (2005), hep-th/0412078.

\bibitem{super2} V.~Cardoso and S.~Yoshida,
  ``Superradiant instabilities of rotating black branes and strings,''
  JHEP {\bf 0507}, 009 (2005), hep-th/0502206.

\bibitem{super3} O.J.C.~Dias,
  ``Superradiant instability of large radius doubly spinning black
  rings,'' Phys.\ Rev.\ D {\bf 73}, 124035 (2006),
  hep-th/0602064.

\bibitem{press} W.H.~Press and S.A.~Teukolsky, ``Floating
Orbits, super-radiant scattering and the black-hole bomb," Nature
{\bf 238}, 211 (1972); \\
V.~Cardoso, O.J.C.~Dias, J.P.S.~Lemos and S.~Yoshida,
  ``The black hole bomb and superradiant instabilities,''
  Phys.\ Rev.\ D {\bf 70}, 044039 (2004)
  [Erratum-ibid.\ D {\bf 70}, 049903 (2004)], hep-th/0404096.

\bibitem{detweiler}
T.~Damour, N.~Deruelle and R.~Ruffini, ``On Quantum Resonances In
Stationary Geometries,'' Lett.\ Nuovo Cim.\
{\bf 15}, 257 (1976);\\
S.~Detweiler, ``Klein-Gordon Equation And Rotating Black Holes,''
Phys.\ Rev.\ D
{\bf 22}, 2323  (1980); \\
H.~Furuhashi and Y.~Nambu,
  ``Instability of massive scalar fields in Kerr-Newman spacetime,''
  Prog.\ Theor.\ Phys.\  {\bf 112}, 983 (2004), gr-qc/0402037.

\bibitem{CardDiasAdS} V.~Cardoso and O.J.C.~Dias,
  ``Small Kerr-anti-de Sitter black holes are unstable,''
  Phys.\ Rev.\ D {\bf 70}, 084011 (2004), hep-th/0405006; \\
  H.~K.~Kunduri, J.~Lucietti and H.~S.~Reall                       
  ``Gravitational perturbations of higher dimensional rotating black holes:
  Tensor Perturbations,''
  Phys.\ Rev.\  D {\bf 74} (2006) 084021, arXiv:hep-th/0606076.



\bibitem{BLMPSV} J.C.~Breckenridge, D.A.~Lowe, R.C.~Myers, A.W.~Peet,
A.~Strominger and C.~Vafa, ``Macroscopic and Microscopic Entropy
of Near-Extremal Spinning Black Holes,'' Phys.\ Lett.\ B {\bf 381}
423 (1996), hep-th/9603078.

\bibitem{finn} M.~Cvetic and F.~Larsen,
``General rotating black holes in string theory: Greybody factors
and  event horizons,'' Phys.\ Rev.\  D {\bf 56}, 4994 (1997),
hep-th/9705192.

\bibitem{CY} M.~Cvetic and D.~Youm, ``General Rotating Five Dimensional Black
Holes of Toroidally Compactified Heterotic String,'' Nucl.\ Phys.\
B {\bf 476}, 118  (1996), hep-th/9603100.

\bibitem{Kthree} M.~Bershadsky, C.~Vafa and V.~Sadov,
``D-Strings on D-Manifolds,''
  Nucl.\ Phys.\  B {\bf 463}, 398 (1996)
  [arXiv:hep-th/9510225];\\
M.B.~Green, J.A.~Harvey and G.W.~Moore,
  ``I-brane inflow and anomalous couplings on D-branes,''
  Class.\ Quant.\ Grav.\  {\bf 14}, 47 (1997)
  [arXiv:hep-th/9605033];\\
Y.K.~Cheung and Z.~Yin, ``Anomalies, branes, and currents,''
  Nucl.\ Phys.\  B {\bf 517}, 69 (1998)
  [arXiv:hep-th/9710206].

\bibitem{cool} C.V.~Johnson and R.C.~Myers,
``The enhancon, black holes, and the second law,''
  Phys.\ Rev.\  D {\bf 64}, 106002 (2001)
  [arXiv:hep-th/0105159].

\bibitem{Breckenridge:1996is}
A.A.~Tseytlin, ``Extreme dyonic black holes in string theory,''
Mod.\ Phys.\ Lett.\ A {\bf 11}, 689  (1996),
hep-th/9601177;\\
J.C.~Breckenridge, R.C.~Myers, A.W.~Peet and C.~Vafa, ``D-branes
and spinning black holes,'' Phys.\ Lett.\ B {\bf 391}, 93 (1997),
hep-th/9602065.

\bibitem{Berti:2005gp}
  E.~Berti, V.~Cardoso and M.~Casals,
  ``Eigenvalues and eigenfunctions of spin-weighted spheroidal harmonics in
  four and higher dimensions,''
  Phys.\ Rev.\ D {\bf 73}, 024013 (2006),
  gr-qc/0511111.

\bibitem{Dias:2007dj}
  O.J.C.~Dias and P.J.~Silva,
  ``Attractors and the quantum statistical relation for extreme (BPS or not)
  black holes,''
  arXiv:0704.1405 [hep-th].

\bibitem{Bardeen:1972fi}
 J.M.~Bardeen, W.H.~Press and S.A.~Teukolsky,
 ``Rotating Black Holes: Locally Nonrotating Frames, Energy Extraction, And
 Scalar Synchrotron Radiation,''
  Astrophys.\ J.\  {\bf 178}, 347 (1972).

\bibitem{DES2007} O.J.C. Dias, R. Emparan and A. Maccarrone, in preparation (2007).

\bibitem{whitehouse} N.~Seiberg and E.~Witten,
``The D1/D5 system and singular CFT,''
  JHEP {\bf 9904}, 017 (1999)
  [arXiv:hep-th/9903224].

\bibitem{skalar} see, for example:\\
J.~Maharana and J.~H.~Schwarz,
  ``Noncompact symmetries in string theory,''
  Nucl.\ Phys.\  B {\bf 390}, 3 (1993)
  [arXiv:hep-th/9207016];\\
A.~Sen, ``String String Duality Conjecture In Six-Dimensions And
Charged Solitonic Strings,''
  Nucl.\ Phys.\  B {\bf 450}, 103 (1995)
  [arXiv:hep-th/9504027].

\bibitem{odd} N.~Kaloper and R.C.~Myers,
  ``The O(dd) story of massive supergravity,''
  JHEP {\bf 9905}, 010 (1999)
  [arXiv:hep-th/9901045];\\
J.~Shelton, W.~Taylor and B.~Wecht,
  ``Nongeometric flux compactifications,''
  JHEP {\bf 0510}, 085 (2005)
  [arXiv:hep-th/0508133].

\bibitem{oddrev} M.~Grana,
  ``Flux compactifications in string theory: A comprehensive review,''
  Phys.\ Rept.\  {\bf 423}, 91 (2006)
  [arXiv:hep-th/0509003].

\bibitem{class} J.M.~Maldacena and A.~Strominger, ``Black hole
greybody factors and D-brane spectroscopy,''
  Phys.\ Rev.\  D {\bf 55} (1997) 861
  [arXiv:hep-th/9609026].

\bibitem{near} J.C.~Breckenridge, D.A.~Lowe, R.C.~Myers, A.W.~Peet,
 A.~Strominger and C.~Vafa, ``Macroscopic and Microscopic Entropy of
 Near-Extremal Spinning Black Holes,''
  Phys.\ Lett.\  B {\bf 381} (1996) 423
  [arXiv:hep-th/9603078].

 \bibitem{page} D.N.~Page,
``Particle Emission Rates From A Black Hole. 2. Massless Particles
From A Rotating Hole,'' Phys.\ Rev.\  D {\bf 14} (1976) 3260.



\end{thebibliography}
\end{document}